\documentclass[aps,prc,preprint,showpacs,nofootinbib]{revtex4-1}
\usepackage{graphics}
\usepackage{epsfig}
\newcommand\nn{\nonumber}
\newcommand\ba{\begin{eqnarray}}
\newcommand\ea{\end{eqnarray}}
\newcommand\be{\begin{equation}}
\newcommand\ee{\end{equation}}

\newcommand{\M} {{\cal M}} 
\begin{document}
\title{Hard light meson production in (anti)proton-hadron collisions and charge-echange reactions}
\author{E.~A.~Kuraev}
\affiliation{\it JINR-BLTP, 141980 Dubna, Moscow region, Russian Federation}
\author{E.~S.~Kokoulina}
\affiliation{\it JINR-VBLHE, 141980 Dubna, Moscow region, Russian Federation}
\author{E.~Tomasi-Gustafsson}
\email{etomasi@cea.fr}
\affiliation{\it CEA,IRFU,SPhN, Saclay, 91191 Gif-sur-Yvette Cedex, France, and
\\ CNRS/IN2P3, Institut de Physique Nucl\'eaire, UMR 8608, 91405 Orsay, France}
\date{\today}
\begin{abstract}
An extension of the QED 'return to resonance' mechanism to light meson emission ($\pi$, $\rho$)
in (anti)proton collisions with a hadronic target (nucleon or nucleus) is proposed. The cross section and the multiplicity distributions are calculated.
The collinear emission (along the beam direction) of a charged meson may be used to produce
high energy (anti)neutron beams.  Possible applications at existing and planned facilities are considered.
\end{abstract}
\maketitle
\section{Introduction}
The first indication of charge-exchange (CE) reactions was found in $\pi ^-p$  and in $pp$ scattering from cosmic rays (see \cite{Murzin}) detected with proportional chambers. CE in $pp$ interactions is defined as
$$
p + p \rightarrow  n + \pi ^{+}  + p + N_{\pi },
$$
where $n$ is a neutron and $N_{\pi }$ is the number of created pions.
The ongoing "Thermalization" project \cite{Avd} detects high multiplicity events in pp interactions with 50 GeV/c beam at U-70 (IHEP, Protvino). CE reactions as well can be investigated at  this facility.

In this work we estimate the CE contribution, using the formalism \cite{BFK, Arbuzov:2010zza}. The emission by the initial proton of a charged light meson-$\pi$ or $\rho$-meson in proton-proton(anti-proton) collisions transforms high energy protons (for example in a proton beam) into neutrons. This effect is observed in accelerator physics \cite{Nikitin}.

CE reactions may occur in antiproton beams, too. High energy, high intensity antiproton beams will be available in next future at PANDA \cite{PANDA}. Hard $\pi$ and $\rho$ meson can be detected with high efficiency. Charged meson will be deviated by the 2T magnetic field of the central spectrometer, and (anti)neutrons, which are produced at high rate, could be used as a secondary high energy beam.

In order to describe CE reactions, let us remind the known QED process of emission of a hard real photons by electron (positron) beams at $e^+e^-$ colliders. Such process enhances the cross section when the energy loss from one of the incident particles lowers the total energy up to  the mass of a resonance. This is known as "return to resonance" mechanism. In the case of creation of a narrow resonance this mechanism appears through a radiative tail: it is the characteristic behavior of the cross section which gradually decreases for energies exceeding the resonance mass. This mechanism provides, indeed, an effective method for studying narrow resonances like $J/\Psi$.

For the emission in a narrow cone along the directions of the initial (final) particles, the emission probability has a logarithmic enhancement, which increases with the energy of the "parent" charged particle. In frame of QED this mechanism is called as "quasi-real electron" mechanism (QRE) \cite{BFK}.

In this work we apply the QRE mechanism to the case of hadrons and, in particular, to the collinear emission of a light meson from a (anti)proton beam. We evaluate the cross section for this process for single as well as multi pion production, where pions can be neutral or charged.

The plan of the paper is as follows. In Section A we recall the formalism of QRE hard photon emission, and then (section B) we extend it to hadronic reactions and give numerical estimations of the cross section for the simplest CE processes. Discussion and conclusions follow.

\subsection{Quasi-real electron kinematics with hard photon emission}

\begin{figure}
\begin{center}
\includegraphics[width=8cm]{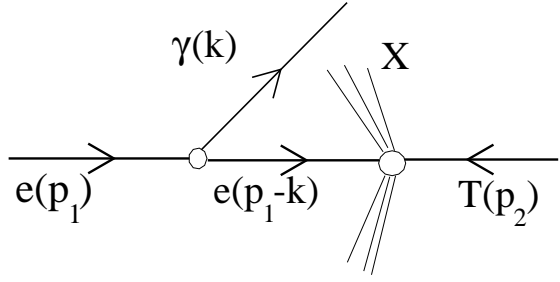}
\caption{Feynman diagram for collinear hard photon emission in $e T$ reactions (T stands for any target)}
\label{Fig:Fig1}
\end{center}
\end{figure}

Let us consider the radiative process  $e(p_1)+T(p_2) \to e(p_1-k) +\gamma (k) + X$ (the four momenta of the particles are written in parenthesis), $T$ stay for any nuclear target(proton $p$, or nucleus $A$), and the final state $X$ is undetected, Fig. \ref{Fig:Fig1}.

The virtual electron after the hard (collinear) photon emission is almost on mass shell \cite{BFK}. This property allows to express the matrix element of the radiative process $e(p)+T \to e(p-k) +\gamma(k)+X$ in terms of the matrix element of the non-radiative process $e+T \to e+X$:
\be
\M_\gamma(p_1,p_2)=e\bar{T}(p_2)\frac{\hat{p_1}-\hat{k}+m}{-2p_1k}\hat{\varepsilon}(k)u(p_1).
\ee

In the case when the denominator of the intermediate electron's Green function is small
 $|(p_1-k)^2-m^2|\ll2p_1p_2$ one can write $\hat{p}_1-\hat{k}+m=\sum_s u^s(p_1-k)\bar{u}^s(p_1-k)$ and the matrix element has a factorized form.

The square of the matrix element, summed over the spin states of the photon is:
\be
\sum |\M_\gamma|^2=e^2\left [\frac{E_p^2+E_{\vec{p}-\vec{k}}^2}{\omega(E_p-\omega)(k p)}-\frac{m^2}{(k p)^2}\right ]\sum|\bar{T}(p_2)u(p_1-k)|^2.
\ee
where $\sum|\bar{T}(p_2)u(p_1-k)|^2$ is the Born matrix element squared with shifted argument.

In the case of unpolarized particles, the cross section of process $e(p_1)+T(p_2)\to e+\gamma+T$ may be written in factorized form:
\ba
d\sigma_\gamma(s,x)&=&d\sigma(\bar{x}s)dW_\gamma(x), s=(p_1+p_2)^2, \bar{x}=1-x, \nn \\
dW_\gamma(x)&=&\frac{\alpha}{\pi}\frac{d x}{x}\left [(1-x+\frac{1}{2}x^2)\ln\frac{E^2\theta_0^2}{m_e^2}-(1-x)\right ],~ x=\frac{\omega}{E}, ~\theta<\theta_0\ll 1, ~\frac{E\theta_0}{m_e}\gg1,
\ea
where $E$ is the energy of the initial electron (center of mass frame implied $\vec{p}_1+\vec{p}_2=0$).

It is assumed here that the initial electron transforms into an electron with energy fraction $1-x$ and a hard photon with
energy fraction $x$ which is emitted within the cone $\theta<\theta_0$ along the direction of initial electron.
Moreover it is implied that $\bar{x}s>s_{thr}$, where $s_{thr}$ is the threshold energy of process without photon emission.
The logarithmic enhancement originates from the small values of the mass of the intermediate electron, which is almost on mass shell. This justifies the name of Quasi Real Electron (QRE) method.

Below we consider a possible extension of the QRE method to the processes with "quasi real" (anti)nucleon intermediate state.

\subsection{Application to hadron physics}

Let us apply this formalism to the case of initial high energy proton (anti-proton) beams and the emission of a hard pion or vector meson in forward direction, collinear to the beam.

\begin{figure}
\begin{center}
\includegraphics[width=8cm]{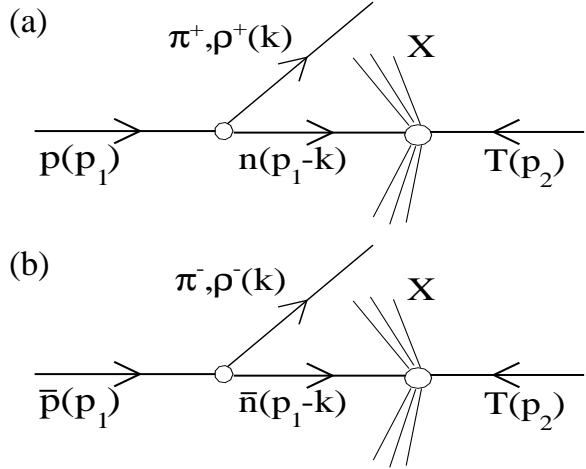}
\caption{Feynman diagram for collinear hard charged pion ($\rho$-meson) emission in $p(\bar p)+T$ collisions.}
\label{Fig:Fig2}
\end{center}
\end{figure}

For the case of emission of a positive-charged $\rho^+$ or $\pi^+$ meson by the high energy (anti)proton, the final state consists in a high energy (anti)neutron, accompanied by a positively charged meson. The charged meson can be deflected by an external magnetic field, providing the possibility to select a high energy neutron beam. In the case of emission of the neutral meson, it can be identified measuring its decay channels.

Let us consider the reactions (Fig. \ref{Fig:Fig2}):
\ba
& p + T&\to n + T +h^+ \\
&\bar p + T& \to \bar n + T +h^-
\ea
where $h=\rho$ or $\pi$ and $T$ may be any target ($p$, $n$, nucleus..). The matrix element for collinear $\pi (\rho)$ emission can be written as:
\ba
\M_{pT}^{h_+}(p_1,p_2))&=& \M_{nT}(p_1-k,p_2)T^{pn}_{h_+}(p_1,p_1-k), \nn\\
\M_{\bar{p}T}^{h_-}(p_1,p_2))&=& \M_{\bar{n}T}(p_1-k,p_2)T^{\bar{p}\bar{n}}_{h_-}(p_1,p_1-k), \nn
\ea
with
\ba
T_{\pi}^{pn}=\frac{g}{m_h^2-2p_1k}\bar u_n(p_1-k)\gamma_5u_p(p_1),~T_{\rho}^{pn}&=&\frac{g}{m_h^2-2p_1k}\bar u_n(p_1-k)\hat\epsilon u_p(p_1).
\ea
The relevant cross sections are:
\ba
d\sigma^{pT\to h_+X}(s,x)&=&\sigma^{nT\to X}(\bar{x}s)d W^{h_+}(x), \nn \\
d\sigma^{\bar{p}T\to h_+X}(s,x)&=&\sigma^{\bar{n}T\to X}(\bar{x}s)d W^{h_-}(x), \nn \\
d\sigma^{pT\to h_0X}(s,x)&=&\sigma^{pT\to X}(\bar{x}s)d W^{h_0}(x).
\ea
The quantity $d W^{\rho}(x)$ can be inferred using the QED result:
\ba
\frac{dW_\rho^i(x)}{dx}&=&\frac{g^2}{4\pi^2}
\frac{1}{x}\sqrt{1-\frac{m_\rho^2}{x^2E^2}}\left [\left (1-x+\frac{1}{2}x^2\right )L-(1-x)\right],
~x=\frac{E_\rho}{E}>\frac{m_\rho}{E}, \nn \\
L&=& \ln\left (1+\frac{E^2\theta_0^2}{M^2}\right ), \rho^i=\rho^+,\rho^-,\rho^0,
\label{eq:eqrho}
\ea
where $M$, $ m_\rho$, $E$, $E_\rho$-are the masses and the energies of the initial proton and the emitted $\rho$-meson (Laboratory reference frame implied).

\begin{figure}
\begin{center}
\includegraphics[width=8cm]{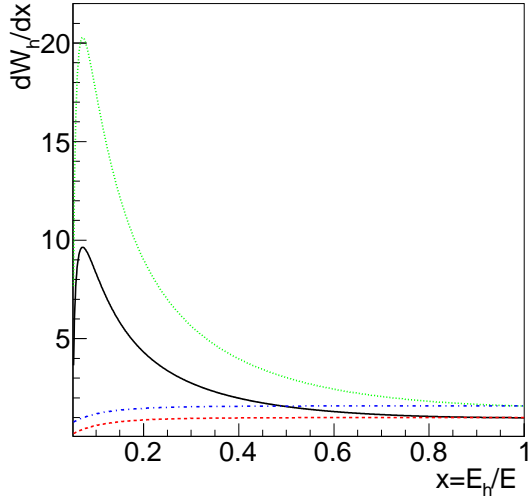}
\caption{Color online. Quantity $dW_h/dx$ for $\rho$- and $\pi$-meson production as a function of the meson energy fraction $x=E_{h}/E$, from Eqs. (\protect\ref{eq:eqrho},\protect\ref{eq:eqpi}), for $E = 15$ GeV and for two values of the $h$-emission angle: $\theta_0=10^{\circ}$ for $\rho$- meson (black, solid) and for $\pi$-meson (red, dashed),
$\theta_0=20^{\circ}$  for $\rho$- meson (green, dotted) and for $\pi$-meson (blue, dash-dotted).}
\label{Fig:hadronprob}
\end{center}
\end{figure}

For the probability of hard pion emission we have
\ba
\frac{d W^\pi}{d x}=\sum|\M_{pn}(p_1,p_1-k)|^2\frac{d^3 k}{16\omega\pi^3},
\ea
with
\ba
\sum|\M_{pn}(p_1,p_1-k)|^2=\frac{g^2}{[m_\pi^2-2(p_1k)]^2}Tr (\hat{p}_1-\hat{k}+M)\gamma_5(\hat{p}_1+M)\gamma_5= \nn \\
\frac{4(p_1k)g^2}{\left [m_\pi^2-2(p_1k)\right ]^2}, (p_1k)=E\omega(1-b c), 1-b^2\approx\frac{m_\pi^2}{\omega^2}+\frac{M^2}{E^2},
\ea
with $c=\cos(\vec{k},\vec{p}_1)$. The angular integration in the region $1-(\theta_0^2/2)<c<1$ leads to
\ba
\frac{dW_\pi^i(x)}{dx}&=&\frac{g^2}{8\pi^2}\sqrt{1-\frac{m_\pi^2}{x^2E^2}}\left [L+\ln\frac{1}{d(x)}+\frac{m_\pi^2}{x d(x)M^2}\right],
\nn \\
x&=&\frac{E_\pi}{E}>\frac{m_\pi}{E},~d(x)=1+\frac{m_\pi^2\bar{x}}{M^2x^2}, ~\bar{x}=1-x,
~\pi^i=\pi^+,\pi^-,\pi^0,
\label{eq:eqpi}
\ea
where  $g=g_{\rho p p}=g_{\pi p p}\approx 6$ is the strong coupling constant.The quantities $d W^h(x)/d x$ as functions of the energy fraction $x=E_h/E$ ($h=\rho,\pi$) are drawn in Fig. \ref{Fig:hadronprob}, for $E=15$, GeV and two different values of $\theta_0$: $10^{\circ}$ and $20^{\circ}$.

The expressions of the integrated probabilities are:
\ba
W_i&=&\int\limits_{x^i_t}^1 \frac{d W_i}{d x}d x=\frac{g^2}{4\pi^2}(A^iL+B^i), \nn\\
A^\rho&=&I_0(x_t^\rho)-I_1(x_t^\rho)+\frac{1}{2}I_2(x_t^\rho),~
B^\rho=-I_0(x_t^\rho)+I_1(x_t^\rho),\nn\\
A^\pi&=&\frac{1}{2}I_1(x_t^\pi);~B^\pi=I_1(x_t^\pi),
\label{eq:eqpr}
\ea
where $x^i_t=E^i_{th}/E$ and $E_{th}$ is the threshold value of the energy of the detected particle, $i=\rho,\pi$,
and the analytic expressions of the functions $I_i(x^i_t)$ for $x_t^i=m^i/E$ are presented in Appendix.

The integrated quantities $W_i$, $i={\rho,\pi}$ can, in general, exceed unity, violating unitarity. To restore unitarity, for the emission of neutral meson, we have to take into account virtual corrections (emission and absorption
of the off-mass shell meson). For this aim we use the known expression for the probability of emission of
$n$ "soft" photons in processes of charged particles hard interaction. The relevant probability
coincides with the Poisson  formula for emission of $n$ soft photons $W_n=(a^n/n!)e^{-a}$ where $a$ is the probability of emission of a
single soft photon \cite{AkhBer81}.

Note that only the emission of "soft" neutral meson is independent and obeys the Poisson distribution. The emission of charged meson evidently,
can not be independent. Fortunately, it is sufficient for our purpose to consider the emission of a charged mesons only at lowest order. This is the reason to introduce a general factor
\be
P_{\pi,\rho}=e^{-W_{\pi,\rho}},
\label{eq:probh}
\ee
which takes into account virtual corrections.

\begin{figure}
\begin{center}
\includegraphics[width=8cm]{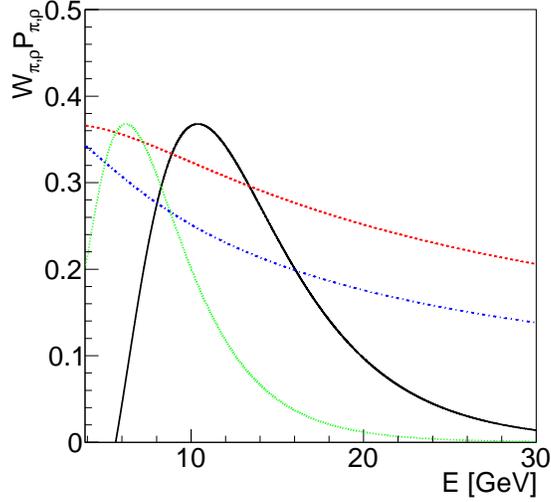}
\caption{Renormalized probabilities $W_{\rho,\pi}\cdot P_{\rho,\pi}$, Eq. (\ref{eq:eqpr}) as function of the incident energy for two values of the hadron emission angle. Notations as in Fig. \ref{Fig:hadronprob}.}
\label{Fig:intprob}
\end{center}
\end{figure}

The renormalized probabilities $W_{\pi,\rho}P_{\pi,\rho} <1$ from Eqs. (\ref{eq:eqpr},\ref{eq:probh}) are illustrated in
Fig. \ref{Fig:intprob} as a function of energy, for two different $\theta_0$ angles.

\begin{figure}
\begin{center}
\includegraphics[width=8cm]{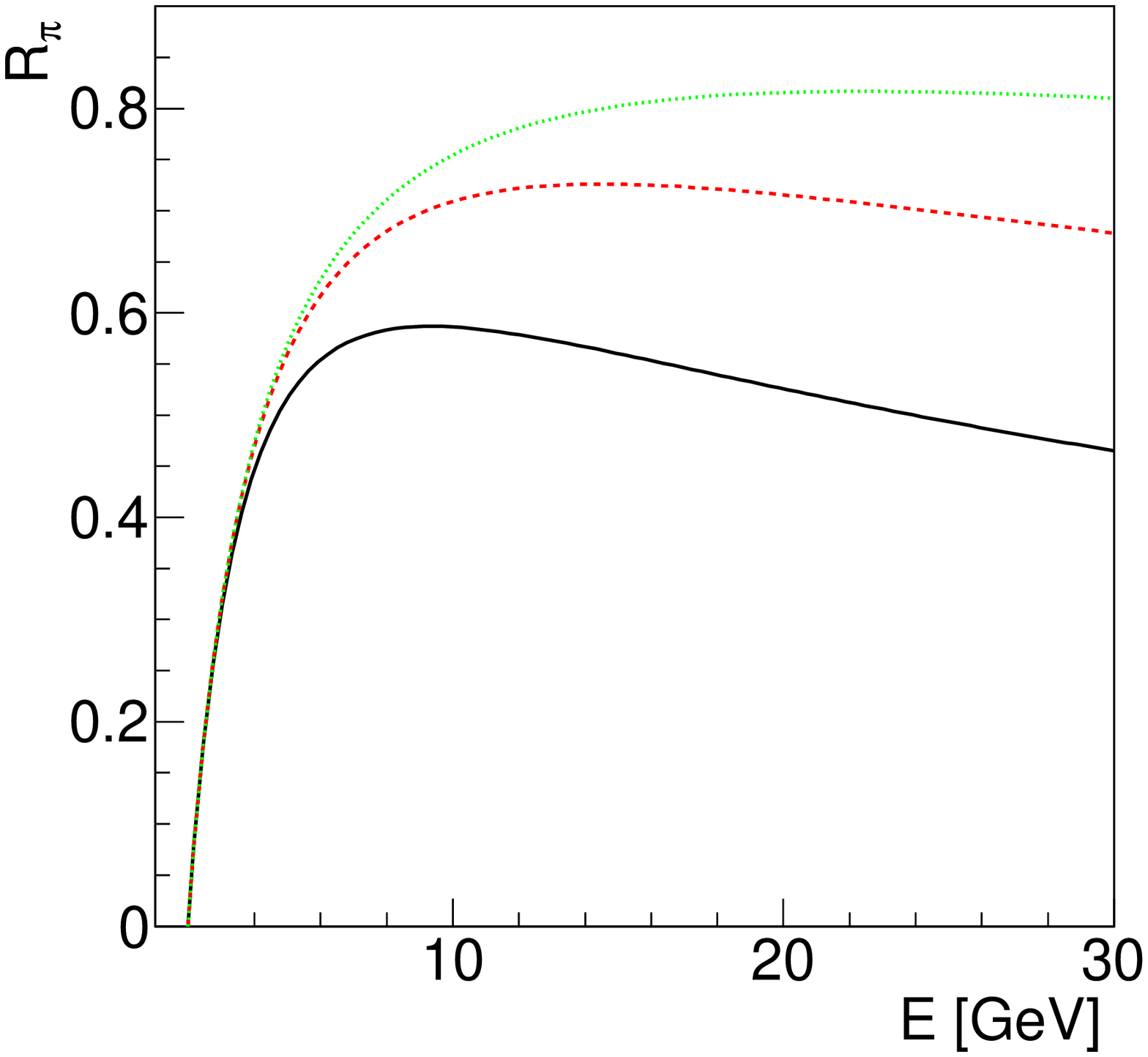}
\caption{Renormalization factor $R_{\pi}$, Eq. (\ref{eq:eqppi}), for the probability of emission of 2 (black, solid line), 3 (red, dashed line), 4 (green, dotted line) pions, as a function of the incident energy for $\theta_0=10^{\circ}$.}
\label{Fig:ppi}
\end{center}
\end{figure}

Keeping in mind the possible processes of emission of $n$ real soft neutral pion escaping the detection, the final result can be obtained using the replacement
\be
\sigma(s) \to \sigma(s)\times {\cal R_{\pi}},~{\cal R_{\pi}}=P_\pi\sum_{k=0}^{k=n}\frac{W^k_\pi}{k!}.
\label{eq:eqppi}
\ee
The renormalization factor ${\cal R}_\pi$ is illustrated in Fig. \ref{Fig:ppi}. for the proability of emission of 2 (black, solid line), 3 (red, dashed line), 4 (green, dotted line) pions.

The quantity $P_\pi$ can be compared to the experimentally measurable phenomena \cite{Nikitin}: the fraction of protons in the final state of
proton-proton collisions is approximately one half, $P_\pi\approx 0.5$. The commonly accepted explanation is that charge exchange reactions are responsible for changing protons into neutrons.


Let us consider the antiproton-proton annihilation into two and three  pions $\bar p+p\to \pi^+\pi^-$ or $\bar p+p\to \pi^+\pi^-\pi^0$.

Concerning the production of two charged pions, accompanied by a final state $X$, we can write:
\be
d\sigma^{p\bar p\to \rho^0 X}= 2\frac {d W_\rho(x)}{dx}\sigma^{p\bar p\to X}(\bar{x}s)\times P_{\rho},~
\label{eq:rhocs}
\ee
where the factor of two takes into account two kinematical situations, corresponding to the emission along each of the initial particles and $P_{\rho}$ is the survival factor (\ref{eq:probh}) which takes into account virtual radiative corrections. The characteristic peak at $x=x_{max}$ has the same nature as for the QED process $e^+ + e^-\to \mu^+ +\mu^- +\gamma$. As explained in Ref. \cite{Baier}, it is a threshold effect, corresponding
to the creation of a muon pair, where $x_{max}=1-4M_\mu^2/s$, $M_{\mu}$ is the muon mass.

\begin{figure}
\begin{center}
\includegraphics[width=8cm]{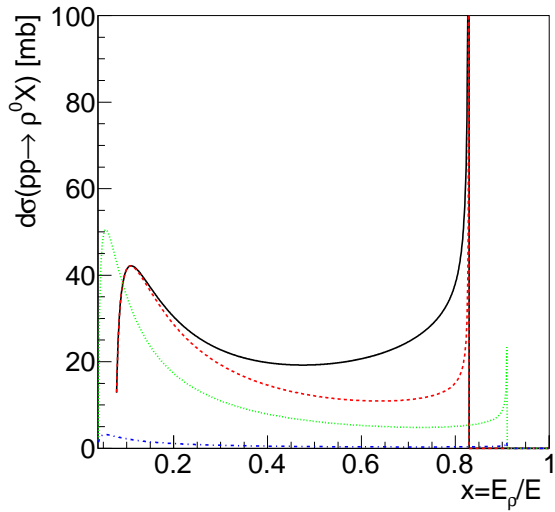}
\caption{The cross section $d\sigma(p,\bar p\to \rho^0 X)$ is plotted as function of the $\rho$ energy fraction for two values of the incident energy and of the $\rho$ emission angle: $E = 10$ GeV and $\theta_0=10^{\circ}$ (black, solid line),   $E = 10$ GeV and $\theta_0=20^{\circ}$ (red, dashed line), $E = 20$ GeV and $\theta_0=10^{\circ}$ (green, dotted line),   $E = 20$ GeV and $\theta_0=20^{\circ}$ (blue, dash-dotted line).}
\label{Fig:cs}
\end{center}
\end{figure}

The cross section (\ref{eq:rhocs}) is illustrated in Fig. \ref{Fig:cs} for two different values of the laboratory energy and of the emitted angle as a function of the $\rho$ meson energy fraction.

In case of three pion production,assuming that the process occurs through a $\pi^0\rho^0$ initial state emission, we find:
\ba
d\sigma(p,\bar{p})^{p\bar{p}\to \pi\rho X}&=&dW^0_\rho(x_\rho) dW^0_\pi(x_\pi)[d\sigma(p-p_\rho,\bar{p}-p_\pi)^{p\bar{p}\to X}+ \nn \\
&&d\sigma(p-p_\pi,\bar{p}-p_\rho)^{p\bar{p}\to X}]P_\pi P_\rho,
\ea
implying the subsequent decay $\rho^0 \to \pi^+\pi^-$.

It is interesting to note that the cross sections for the interaction of high energy neutron (anti-neutron) beams with a hadronic target can be calculated using the cross sections of proton beam interacting with the same target with the emission of the charged meson. We obtain (see Eqs. (\ref{eq:eqrho},\ref{eq:eqpi})):
\ba
\sigma^{nT\to X}(\bar{x}s)=\frac{d\sigma^{pT\to h^+ X}/d x}{d W_+(x)/d x},
\ea
and similarly for the anti-proton beams.

To be definite if we use the experimental data for the total cross section of process $\bar{p}p \to \bar{n} h^-p\approx$ 1mb, (for a compilation, see \cite{Dbeyssi:2012zz}) we can predict the value of the total cross section  of process $\bar{n}p \to X$
\ba
P_\pi W_\pi(E_1,\theta_0)\sigma^{\bar{n}p\to X}(E-E_1) =\sigma^{\bar{p}p\to \pi X}(E),
\ea
with $W_\pi(E,\theta_0)$  given in Eqs. (\ref{eq:eqrho},\ref{eq:eqppi}).

\section{Conclusions}

We have extended the QRE method to light meson emission from an (anti)proton beam. We have calculated the probabilities for multi-pion emission and the relative cross section. The considered processes can be measured at present and planned hadron facilities.

Note that the probabilities to create a $\pi$ or $\rho$-meson by a proton, can also be obtained using the infinite momentum reference frame, (Ref.  \cite{Altarelli:1977zs}, Eq. (52)).

The arguments given above have a phenomenological character and are formulated in terms of hadrons. A similar idea, at quark level, was introduced in Ref. \cite{Teryaev82}, where the emission of $\rho$-meson by quark and the $\rho$ meson production in quark-antiquark annihilation was studied. Special attention was paid to polarization phenomena of the created $\rho$-meson.

We have also suggested a possible application.  The collinear light meson emission could also be used to produce secondary (anti)neutron beams, at a high energy (anti)proton accelerator. This would constitute an alternative to the usual way, when high-energy neutron beams are produced as secondary beams, by break-up of  deuterons on a hadronic target.

In frame of the "Gluon Dominance Model", developed by one of us \cite{GDM} the ratio of the inelastic CE  cross section to the total inelastic cross section in $pp$ scattering is estimated as 40\%, in reasonable agreement with the experimental data \cite{Murzin}.

The collinear light meson emission mechanism in (anti)proton-proton collisions provide a possible source of events with rather high multiplicities of (charged and neutral) pion production. For such events the emission of hadrons in initial as well as in final states must be taken into account.

The simplest CE processes $p\bar{p} \to n \bar{n}$,~ $\bar{p} n \to \bar{\Lambda} \Sigma^-$ can be in principle measured at PANDA. Other reaction mechanisms can contribute to these processes. In case of a description in terms of a single pseudoscalar meson $\pi^+, K^+$ exchange, information on the strange meson-barion constant can be extracted.

\section{Acknowledgements}

One of us $EAK$ is grateful to RFBR grant 11-02-00112 for support. We are grateful to S. Barkanova, A. Alexeev, and V. Zykunov, for interest to this problem and to V.A.~Nikitin for useful discussions.

\section{Appendix}

The analytic expressions for calculating the integrals
$$I_n(z)=\int\limits_z^1\frac{d x}{x}x^n\sqrt{1-\left(\frac{z}{x}\right )^2}$$
are:
\ba
I_0(z)&=&\frac{1}{2}\ln\frac{1+r}{1-r}-r; \nn \\
I_1(z)&=&r+z \arcsin (z); \nn \\
I_2(z)&=&\frac{1}{2}r-\frac{z^2}{4}\ln\frac{1+r}{1-r}; \nn
\ea

with  $r=\sqrt{1-z^2}$.

\end{document}